\documentclass[preprint,12pt,authoryear]{elsarticle}

\usepackage{amssymb}
\usepackage{float}
\usepackage[none]{hyphenat}

\usepackage[inline]{trackchanges}
\addeditor{CH}
\addeditor{MK}

\journal{Ecological Modelling}

\begin{document}

\begin{frontmatter}

\title{Statistical measures of complexity applied to ecological networks}

\author[unrn,irnad]{Claudia A. Huaylla\corref{cor1}}
\ead{cahuaylla@unrn.edu.ar}

\cortext[cor1]{Corresponding author}

\author[cab,ib]{Marcelo N. Kuperman}
\ead{kuperman@cab.cnea.gov.ar}

\author[unrn,irnad]{Lucas A. Garibaldi}
\ead{lgaribaldi@unrn.edu.ar}

\address[unrn]{Universidad Nacional de R\'{\i}o Negro. Instituto de Investigaciones en Recursos Naturales, Agroecolog\'{\i}a y Desarrollo Rural. San Carlos de Bariloche, R\'{\i}o Negro, Argentina.}
\address[irnad]{Consejo Nacional de Investigaciones Cient\'{\i}ficas y T\'{e}cnicas. Instituto de Investigaciones en Recursos Naturales, Agroecolog\'{\i}a y Desarrollo Rural. San Carlos de Bariloche, R\'{\i}o Negro, Argentina.}
\address[cab]{Centro At\'omico Bariloche (CNEA) and CONICET, R8402AGP Bariloche, Argentina.}
\address[ib]{Instituto Balseiro, Universidad Nacional de Cuyo, R8402AGP Bariloche, Argentina.}

\begin{abstract}

Networks are a convenient way to represent many interactions among different entities as they provide an efficient and clear methodology to evaluate and organize  relevant data.
While there are many features for characterizing networks there is a quantity that seems rather elusive: Complexity. The quantification of the complexity of networks is nowadays a fundamental problem. Here, we present a novel tool for identifying  the complexity of ecological networks.
We compare the behavior of two relevant indices of complexity: K-complexity and Single value decomposition (SVD) entropy. For that, we use real data and null models. Both null models consist of randomized networks built by  swapping a controlled number of links of the original ones.
We analyze 23 plant-pollinator and 19 host-parasite networks as case studies. Our results show interesting features in the behavior for the K-complexity and SVD entropy with clear differences between pollinator-plant and host-parasite networks, especially when the degree distribution is not preserved. 
Although SVD entropy has been widely used to characterize network complexity, our analyses show that K-complexity is a more reliable tool. Additionally, we show that degree distribution and density are important drivers of network complexity and should be accounted for in future studies.

\end{abstract}


\end{frontmatter}


\section{Introduction}

Ecologists have turned to network theory to analyze ecological systems due to its ability to provide a powerful mathematical framework to capture the complexity and structure of ecological communities. The use of networks has allowed researchers to identify how the topological structure of ecological systems is linked to their ecological properties and processes. As a result, numerous measures have been proposed to capture different aspects of network structure and its relationship with ecological systems \citep{Delmasetal2019}. Most studies have focused on bipartite networks that describe single interaction types, with mutualistic and antagonistic systems being the most commonly studied \citep{MorrisonDirzo2020}. Researchers have examined various network indices to identify their structural properties. However, the degree of complexity of these networks has not been fully explored, and it is essential to compare and contrast different complexity values \citep{Bascompte2003,Poulin2010}. 

Recently, researchers have become increasingly interested in networks that contain multiple interaction types \citep{Fontaineetal2011, Melian2009, Pilosofetal2017, Huayllaetal2021}, and various indices for bipartite and multilayer networks have been studied \citep{Strydom2021}. A tool has been developed to determine whether a network encodes relevant information in its system. The complexity of networks has been studied using different indices, including the Singular Value Decomposition (SVD) entropy index. However, it is not enough to calculate the initial complexity value of a network; it is also necessary to compare it with other values. This aspect has not been thoroughly explored. While some results derived from network-based studies help to characterize their structure, this research aims to go beyond the traditional paradigm based on networks. The proposed procedure can be applied to all types of networks.

Various indices can be used to approach the complexity of a network, but it is important to consider the type of network and tool used. Complexity can be defined in different ways, such as the product of entropy and disequilibrium or a set of indices such as Singular Value Decomposition (SVD) entropy and K-complexity. When approaching complexity as the product of entropy and disequilibrium, the calculation of entropy must be taken into account, which can vary depending on the defined random variable. For example, using three different random variables can result in three different entropy values.  If we consider a matrix $10\times10$ and its ratio of ones and zeros, the random variable $X=\{x_0=0.7, x_1=0.3\}$ its entropy is  $H(X) = 0.88$. Another way to describe a graph is to consider the degree of each node $D = <3, 3, 3, 3, 3, 3, 3, 3, 3, 3>$ in this situation the entropy is $H(D) = 0$. Other option is consider the degree distribution $PD= \{d_0 = 0, d_1 = 0, d_2 = 0, d_3 = 1\}$  the entropy is  $H(PD) = 0$. Depending on the defined random variable, three different results can be obtained. The Kolmogorov complexity is a more robust and reliable measure that can be approximated in various ways, such as using the probability algorithm presented by Solomonoff and Levin \citep{Levin1974, Solomonoff19641}, or the Block Decomposition Method for larger strings or matrices \citep{Levin1974}. This measure is widely used in areas other than ecology. 

Recently, complexity has also been approached as the decomposition of the original matrix, known as Singular Value Decomposition (SVD) entropy (Strydom et al. 2021). This method uses the entropy measure of Shannon and applies it to the non-zero singular values of the matrix obtained through SVD. This index will be explained in detail in section 2.
However, when considering complexity as a set of indices, it is important to compare them to a null model to avoid missing the whole picture. For instance, connectance and nestedness are some indices that need to be compared with a null model to fully understand their values.

To achieve our study's primary objective, we employ network characterization methods using information gathered from website data. We examine modularity, entropy, SVD entropy, and K-complexity to identify the tools that enable us to distinguish network complexity, by comparing their structures to randomized versions. We present two systematic approaches to gradually increase the randomness in a network and demonstrate how they affect modularity, entropy, SVD entropy, and K-complexity. Two algorithms that generate random networks were utilized for this purpose. Our research objectives are (1) to determine the complexity level and (2) to identify the most appropriate tools based on the network we're characterizing. Section 2 includes the details on the algorithm implementation, tool identification, and dataset, while our results and interpretations are provided in Section 3. In Section 4, we present our discussion and the implications of our findings.

\section{Materials and Methods}

\subsection{Statistical measures frequently used in network characterization}

To assess the complexity of ecological networks, we analyzed 23 plant-pollinator networks and 19 host-parasite networks from the web-of-life.es database. This database includes species interaction networks compiled from supplementary materials from various habitats, organisms, sampling years, and methodologies.

Different parameters can be used to characterize a network, such as the number of links, nodes, assortativity, nestedness, degree distribution, density, modularity, and entropy. In our study, we focused on calculating the number of links and nodes, assortativity, nestedness, density, modularity, and entropy. The number of nodes represents the number of species in the network, and the number of links denotes the interactions between species. Assortativity coefficient quantifies the homophyly level in the graph, based on the degree of each vertex. Nestedness is the property that, for any two nodes $i$ and $j$, if the degree of $i$ is smaller than the degree of $j$, then the neighborhood of $i$ is contained in the neighborhood of $j$. Density is the ratio of actual links to possible links in a network. Modularity measures the intensity of intra-community versus inter-community links, and it helps in characterizing the community structure. Communities refer to disjoint sets of nodes that share common properties and/or roles within the network. 

Particularly, modularity can be calculated by using different algorithms \citep{Girvan2002, Newman2004a, Newman2004b, Blondeletal2008, Barber2007}. The Louvain  algorithm, that  was used in this work, groups the nodes in such a way as to maximize the modularity function, considers several groups of nodes, and selects the groups that maximize the function. 
The entropy is calculated using a random variable. In this paper, we have defined a discrete random variable taking $K$ values. The clusterings were calculated using the Louvain algorithm. For each clustering, we calculated $P(k)=\frac{n_k}{n}$ where $n_k$= number of nodes in that cluster and $n=$ number of total nodes in a network, $k=$ number of clusters. The entropy was defined as (Meila et al. 2007) 
\begin{equation}
     h(c)=-\sum_{k=1}^K  P(k) \cdot log(P(k))
\end{equation}

\subsection{Tools for estimating the complexity of a network: Singular Value Decomposition and Kolmogorov complexity}

There are many tools for estimating the complexity of a network, in this study two tools are presented, the singular value decomposition (SVD) entropy and the Kolmogorov complexity (K-complexity). The first one is the factorization of a matrix A (where $A_{m,n} \in \mathbb{R}$) into $U \cdot  D \cdot V^T$.  $U \in \mathbb{R}^{m \times m}$  is an orthogonal matrix and $V \in \mathbb{R}^ {n \times n}$ is an orthogonal matrix. 
$D \in \mathbb{R}^{m \times n}$ is a matrix that only contains non-negative $d$ values along
its diagonal and all other entries are zero. The SVD entropy is calculated using $s_i=\frac{d_i}{\sum_i(d_i)}$, where $d_{i}=D_{ii}$ are the diagonal elements, these values are known as the singular values of A. Then J is calculated
\begin{equation}
J=-\frac{1}{ln(k)}\sum_{i=1}^k s_i ln(s_i), \end{equation} where $k = rk(A)$   \citep{Strydom2021}.

The Kolmogorov complexity (K-complexity) is much more reliable and robust, however, it is incomputable. Fortunately, K-complexity can be approximated using different algorithms. One way for approximating the true value of K is to use the algorithmic probability introduced
by Solomonoff and Levin \citep{Levin1974, Solomonoff19641}. The algorithmic probability used to calculate K-complexity is the Coding Theorem (CTM) \citep{Levin1974}. The problem is that the CTM can be applied only to short strings consisting of 12 characters or less. For larger strings and matrices, the BDM (Block Decomposition Method) should be used. The
BDM use the decomposition of the string $s$  into (possibly overlapping) blocks $\{b_1,b_2, \cdots, b_k\}$ \citep{Zeniletal2016, Morzy2017}.

\subsection{Null models}
In this study, we employ two null models to create randomized versions of the original networks. The first null model preserves the degree distribution of each node, while the second null model only maintains the total number of links. To achieve this, in the first null model, we randomly select two pairs of connected nodes and exchange their links to create two pairs of connected nodes that differ from the original ones. This procedure is elaborated in \cite{Huayllaetal2021}. In the second null model, we select a single pair of connected nodes and disconnect one of the nodes at one end of the link. We then connect the free end to another node that is randomly selected. This approach is based on Watts and Strogatz (1998) \citep{Watts1998, kuperman2001}. In both null models, we take special care to avoid double links and disconnected nodes.

\begin{figure}[H]
\centering
\includegraphics[scale=0.3]{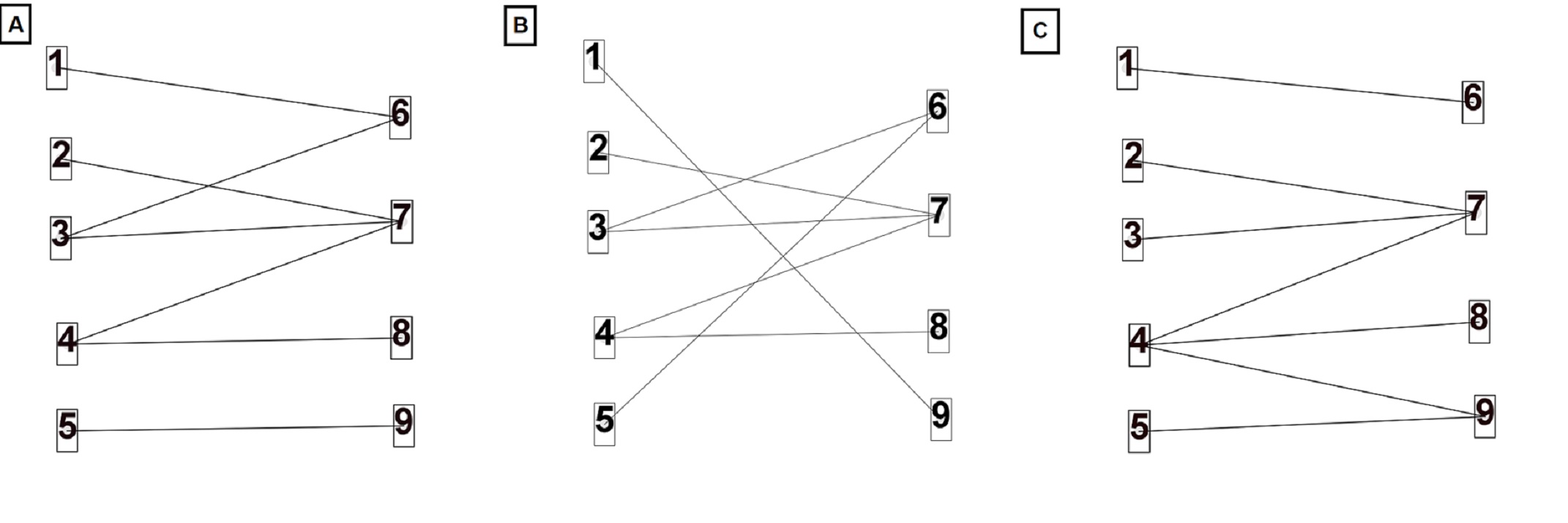}
\caption{Simplified representation of the swap procedure. (A) Original graph. $A_{16}=1$, $A_{59}=1$, $A_{56}=0$, $A_{19}=0$, $D_4=2$, $D_6=2$ . (B) Graph with a single swap. $A_{16}=0$, $A_{59}=0$, $A_{56}=1$, $A_{19}=1$, the degree distribution is the same as the original graph. (C) Graph with a single swap. $D_4=3$, $D_6=1$.}
\label{modelosvarios}
\end{figure}

\subsection{Comparing measures}

To assess the efficacy of the protocol, we utilized 42 bipartite networks, consisting of 19 host-parasite and 23 plant-pollinator networks. These datasets were obtained from the web-of-life.es website, which features a wide variety of species interactions from various continents, environments, and methodologies over numerous sampling years.

Simply calculating the initial values of entropy, modularity, SVD entropy, and K-complexity for each network is inadequate; it is essential to compare them with other values for the same network. If the original network is complex, the measured parameters should deviate from those of a random network. Therefore, we introduced sequential and controlled topological changes to networks to increase disorder and disrupt existing structures in order to transition them to their randomized versions. By tracking these indices through these changes, their values should shift from the initial ones to those of a random network with similar topologies. Our approach involved two models: the first preserved the degree distribution, while the second maintained the number of links while altering the degree distribution.

To compare both groups of networks after the changes in each network, we plotted SVD entropy versus changes, then fitted those points using a linear model to identify the sign of the main coefficient ($\beta_{svd}$). The same was done for K-complexity versus changes ($\beta_{k}$). We are interested in the behavior after the changes. For each network and for each model, two main coefficients were calculated, $\beta_{svd}$ and  $\beta_{k}$. Finally, we compared these measures.

\begin{figure}[H]
\centering
\includegraphics[scale=0.45]{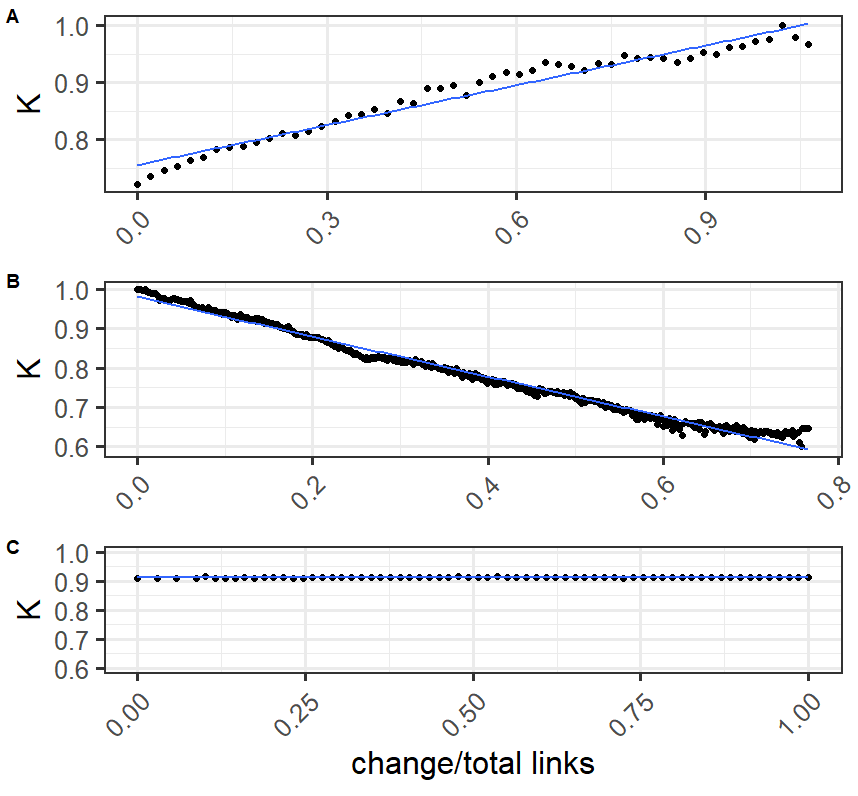}
\caption{A) $\beta_k > 0$. K-complexity vs. changes/total links in the original network. Each value is a 10 network average except for the original one. B) $\beta_k < 0$. K-complexity vs. changes/total links (swappings/total links) in the original network. Each value is a 10 network average except for the original one. C) $\beta_k \approx 0$. K-complexity vs. changes/total links (swappings/total links) in the original network. Each value is a 10 network average except for the original one.}
\label{pendientes}
\end{figure}

\section{Results}

\subsection{Statistical measures}

We have calculated the number of links, number of nodes,  assortativity, nestedness, and density of each network. These results can be found in the appendix. The number of nodes in the networks go from 27 to 997 and the number of links from 32 to 2933. The nestedness was calculated revealing that  the networks of pollinator-plant are more nested than host-parasite networks (see appendix. fig A 1). The density in these networks is less than 0.3 and these networks have negative assortativity. 

\subsection{Null models}

\subsubsection{Preserving the degree distribution}

The original networks were randomized and, preserving the degree of each node, the entropy, modularity, SVD entropy and K-complexity  were calculated. We calculated the $\beta_{entropy}, \beta_{modularity}, \beta_{SVD}, \beta_{k}$ generated in each situation to identify which indices increased or decreased as we made changes.

The results showed that for pollinator-plant networks, modularity decreases and entropy increases  as the number of changes increases, these can be observed with the sign of $\beta_{modularity}$ and $\beta_{entropy}$ (Fig. 3.C-Fig 3.D). Furthermore, the disorder of the network grew as the entropy values moved from the original ones to the corresponding random ones. The modularity and entropy showed in a complementary way that pollinator-plant networks have some coded information that was destroyed when randomness was incorporated. However, in the case of host-parasite networks the disorder is not reflected using modularity and entropy. We observed that sometimes $\beta_{modularity}>0$ and other times $\beta_{modularity}<0$.  The density of the host-parasite networks is higher than the pollinator-plant one. The density is important and may be the reason why certain tools do not reflect what is expected. 

In this study, the pollinator-plant networks have a density of less than 0.2. If networks have a density greater than 0.15, entropy may not be an indicator to measure disorder. The entropy was calculated using the communities but it does not reflect the change that is originally made, as certain nodes will not change communities because they do not have much margin for change.

Other indices were calculated to determine the complexity of the networks. SVD entropy showed no significant change in the case of pollinator-plant networks as we make changes and almost no variation, the variation was between 0 and 0.1.
In the case of host-parasite networks, we observed that for some networks there was a more pronounced change than in other cases.

K-complexity showed that the pollinator-plant networks are more complex and that they lose complexity as the level of randomization increases. $\beta_k$  generally takes a negative value and the host-parasite network generally takes a positive value. This means that in the case of pollinator-plant networks the values decreased when the changes increase.   

\begin{figure}[H]
\centering
\includegraphics[scale=0.43]{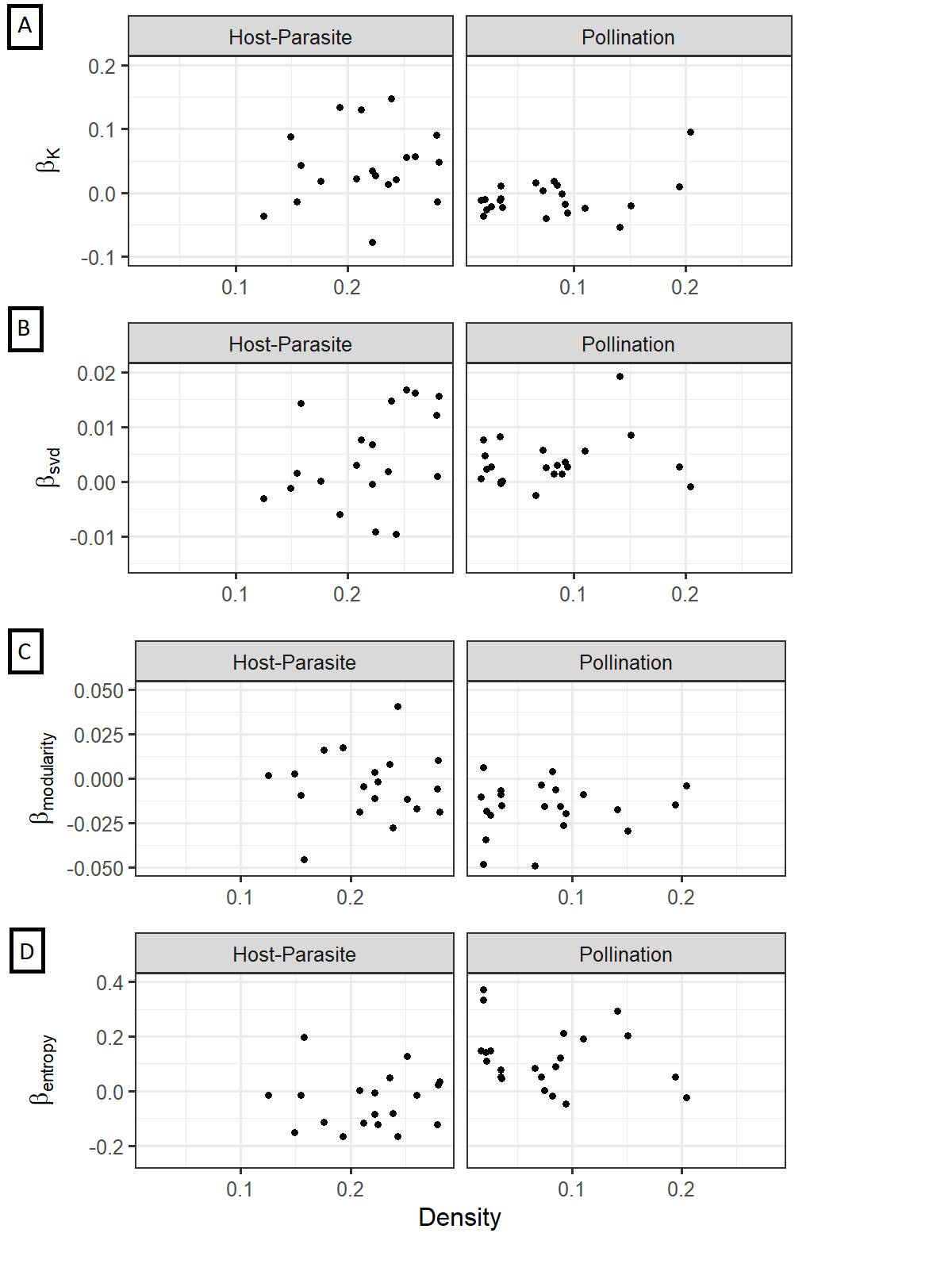}
\caption{A) Relation between $\beta_k$ and density for each network. B) Relation between $\beta_{svd}$ and density for each network. C) Relation between $\beta_{modularity}$. D) Relation between $\beta_{entropy}$}
\label{modelo1}
\end{figure}

\subsubsection{Preserving the number of links}

Using the second method, which preserves the number of links but changes their degree distribution, SVD entropy, and K-complexity were calculated as we made changes to determine the degree of complexity of the host-parasite networks. It is noted that in the pollinator-plant networks, the value of K-complexity decreases as we make the changes, but the same does not happen when we use the host-parasite networks, i.e. these networks do not present complexity in their structure. Figure 4.A shows that in most of the pollinator-plant networks, $\beta_k$ is negative when we make the changes unlike in the host-parasite networks.
Using SVD entropy we saw that, when the changes increase they become more complex networks, this may be because their density is higher than 0.1. SVD entropy is still not a reliable tool to determine the complexity of a network.
$\beta_{entropy} >0$ for the pollinator-plant networks, which means that the original network has information in its system but for host-parasite networks sometimes $\beta_{entropy}$ is negative and other times is positive, this means that these networks have not information in their system. 
The degree distribution was changed and the modularity was influenced by this change. Figure 4.C shows that the values of modularity, in general, are greater than zero because when the degree distribution changed the cluster too and this produced remarkable changes in this index.

\begin{figure}[H]
\centering
\includegraphics[scale=0.43]{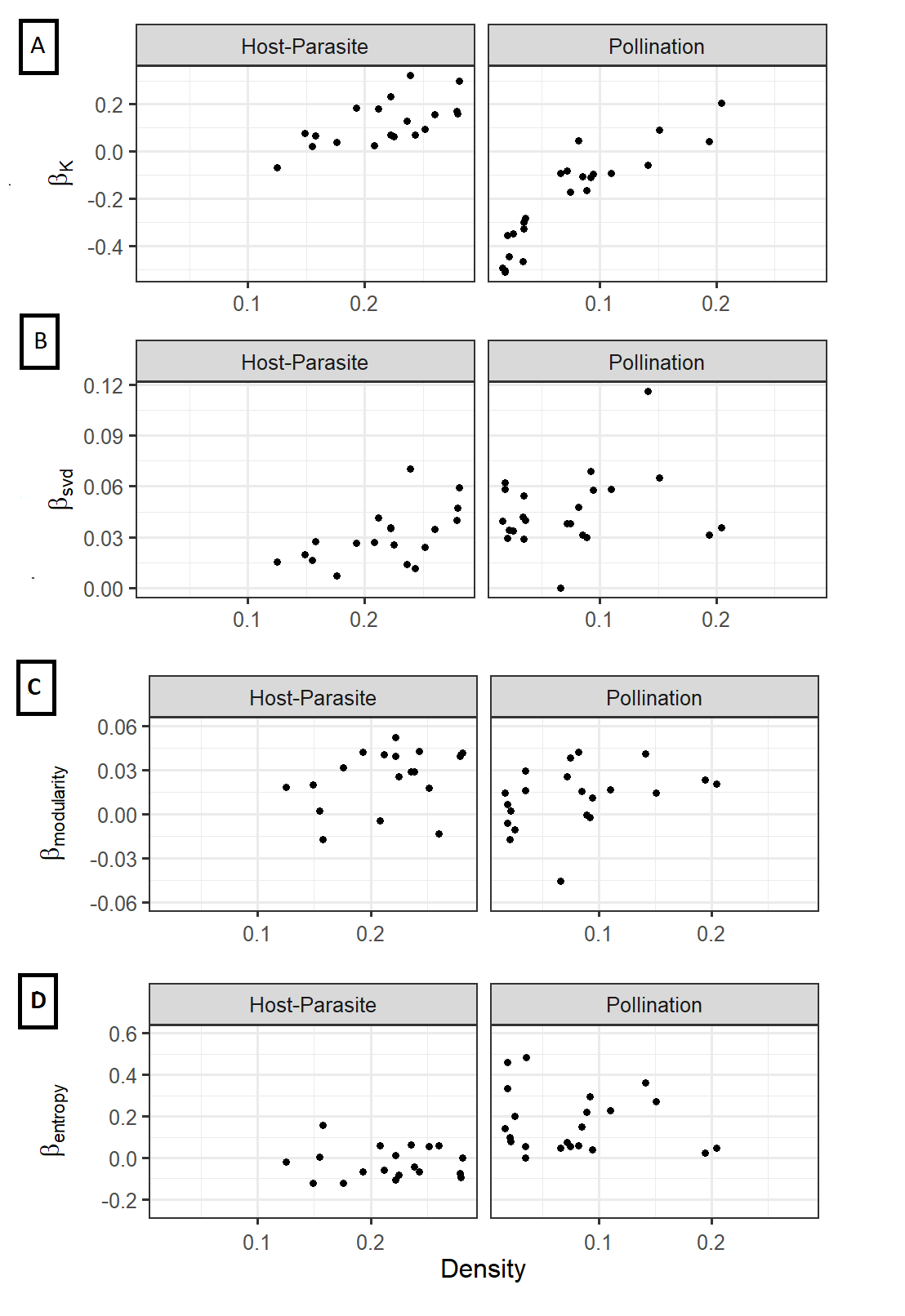}
\caption{A) Relation between $\beta_k$ and density for each network. B) Relation between $\beta_{svd}$ and density for each network. C) Relation between $\beta_{modularity}$. D) Relation between $\beta_{entropy}$}
\label{modelo2}
\end{figure}

It is worth noting that the degree distribution has system information, so that information is lost when changes are made to the network, as can be seen with plant-pollinator networks. The second method makes it clear that the degree distribution is relevant in network analysis.

It is reasonable that the value decreases in plant-pollinator networks because the original network has a high initial complexity and when we randomize the network it loses its complexity.
It is worth noting that the subset of networks selected for the study reflects what was already known, that pollinator networks are more nested than parasitoid networks, but it was nevertheless possible to make progress on the complexity of a network. 

The initial values of SVD entropy were calculated (see appendix fig A 2), although the values were high, this does not mean that the networks are complex, i.e. it is not enough to calculate the initial values of the original networks, we must randomize them to determine if the networks present a degree of complexity. 

\section{Discussion}

The present study aims to characterize different aspects of the complexity of ecological networks. We used 23 networks of plant-pollinator and 19 networks of host-parasite interactions from the web-of-life.es database. The following parameters were calculated to characterize a network: number of links and nodes, assortativity, nestedness, density, modularity, and entropy. The study also used two tools to estimate network complexity: singular value decomposition (SVD) entropy and Kolmogorov complexity (K-complexity). The SVD entropy quantifies the factorization of a matrix into orthogonal matrices, while the K-complexity approximates the true value of complexity using different algorithms. 

Additionally, the study used two null models, one that preserves the degree of each node and another that preserves the total number of links. The former randomly exchanges pairs of connected nodes, while the latter disconnects a single node and reconnects it to another node. Despite the increasing randomness, some features of the network, such as the degree distribution and number of links, remain constant. Thus, we evaluated the changes in the original network as we added more randomness while preserving either the degree distribution or the number of links. We found that K-complexity decreased monotonically until a completely random network was obtained for the pollinator-plant network with the constraint to preserve the number of links. Additionally, we discovered that entropy and modularity are reliable tools for networks with a density lower than 0.1.

Our work differs from previous studies that fundamentally analyze some characteristics of bipartite networks because we determine in advance whether a network is complex before working on it. Furthermore, we selected a set of tools to characterize them. Entropy and modularity are useful tools to determine if a network contains relevant information in its system. As changes increase, the entropy value follows an increasing trend, indicating the increasing disorder and loss of information from the original network. Modularity complements the results obtained by entropy by showing that the community's structure is being lost. When this does not occur, we can use other tools to determine if the network contains relevant information.

As mentioned above, We  choose between two null models. If the density is less than 0.1, the first model will be used. In these cases, entropy and modularity are sufficient to determine if the network has a structure different from that of random networks. If the density is greater than 0.1 and we do not obtain results on the original network, we will use the second model, which preserves the number of links but changes the degree distribution of the original network. Using the second model allows us to calculate other indices, such as SVD entropy and K-complexity, to evaluate the complexity of a network. The algorithms and indices we selected can be applied to networks from different areas.

Although we only considered bipartite networks in this study, our method can be applied to other types of networks. This approach can be useful to identify when a network is not complex or does not have information. If a network lacks information, it may be because some of its links should not exist, leading us to investigate the accuracy of its current links. Additionally, by studying networks without complexity in their systems, we can predict future links between nodes. Link prediction is a growing area that can help identify irrelevant links in a network. Finally, our algorithms and selected indices can be applied to a wide range of disciplines to verify whether a network contains valuable information and its level of complexity.

 \bibliographystyle{elsarticle-harv} 
 \newpage
 \bibliography{cas-refs}

\begin{thebibliography}{20}
\expandafter\ifx\csname natexlab\endcsname\relax\def\natexlab#1{#1}\fi
\providecommand{\url}[1]{\texttt{#1}}
\providecommand{\href}[2]{#2}
\providecommand{\path}[1]{#1}
\providecommand{\DOIprefix}{doi:}
\providecommand{\ArXivprefix}{arXiv:}
\providecommand{\URLprefix}{URL: }
\providecommand{\Pubmedprefix}{pmid:}
\providecommand{\doi}[1]{\href{http://dx.doi.org/#1}{\path{#1}}}
\providecommand{\Pubmed}[1]{\href{pmid:#1}{\path{#1}}}
\providecommand{\bibinfo}[2]{#2}
\ifx\xfnm\relax \def\xfnm[#1]{\unskip,\space#1}\fi
\bibitem[{Barber(2007)}]{Barber2007}
\bibinfo{author}{Barber, M.J.}, \bibinfo{year}{2007}.
\newblock \bibinfo{title}{Modularity and community detection in bipartite
  networks}.
\newblock \bibinfo{journal}{Phys. Rev. E} \bibinfo{volume}{76},
  \bibinfo{pages}{066102}.
\newblock \URLprefix \url{https://link.aps.org/doi/10.1103/PhysRevE.76.066102},
  \DOIprefix\doi{10.1103/PhysRevE.76.066102}.
\bibitem[{Bascompte et~al.(2003)Bascompte, Jordano, Meli\'an and
  Olesen}]{Bascompte2003}
\bibinfo{author}{Bascompte, J.}, \bibinfo{author}{Jordano, P.},
  \bibinfo{author}{Meli\'an, C.}, \bibinfo{author}{Olesen, J.},
  \bibinfo{year}{2003}.
\newblock \bibinfo{title}{The nested assembly of plant-animal mutualistic
  networks}.
\newblock \bibinfo{journal}{Proc. Natl. Acad. Sci. USA 100}
  \bibinfo{volume}{16}, \bibinfo{pages}{9383–9387}.
\newblock \DOIprefix\doi{10.1073/pnas.1633576100}.
\bibitem[{Blondel et~al.(2008)Blondel, Guillaume, Lambiotte and
  Lefebvre}]{Blondeletal2008}
\bibinfo{author}{Blondel, V.D.}, \bibinfo{author}{Guillaume, J.L.},
  \bibinfo{author}{Lambiotte, R.}, \bibinfo{author}{Lefebvre, E.},
  \bibinfo{year}{2008}.
\newblock \bibinfo{title}{Fast unfolding of communities in large networks}.
\newblock \bibinfo{journal}{J. Stat. Mech.-Theory Exp.} \bibinfo{volume}{10},
  \bibinfo{pages}{P10008}.
\newblock \URLprefix \url{http://dx.doi.org/10.1088/1742-5468/2008/10/P10008},
  \DOIprefix\doi{10.1088/1742-5468/2008/10/P10008}.
\bibitem[{Delmas et~al.(2019)Delmas, Besson, Brice, Burkle, Dalla~Riva, Fortin,
  Gravel, Guimarães, Hembry, Newman, Olesen, Pires, Yeakel and
  Poisot}]{Delmasetal2019}
\bibinfo{author}{Delmas, E.}, \bibinfo{author}{Besson, M.},
  \bibinfo{author}{Brice, M.}, \bibinfo{author}{Burkle, L.},
  \bibinfo{author}{Dalla~Riva, G.}, \bibinfo{author}{Fortin, M.},
  \bibinfo{author}{Gravel, D.}, \bibinfo{author}{Guimarães, P.J.},
  \bibinfo{author}{Hembry, D.}, \bibinfo{author}{Newman, E.},
  \bibinfo{author}{Olesen, J.}, \bibinfo{author}{Pires, M.},
  \bibinfo{author}{Yeakel, J.}, \bibinfo{author}{Poisot, T.},
  \bibinfo{year}{2019}.
\newblock \bibinfo{title}{Analysing ecological networks of species
  interactions}.
\newblock \bibinfo{journal}{Biol Rev.} \bibinfo{volume}{94},
  \bibinfo{pages}{16--36}.
\newblock \DOIprefix\doi{10.1111/brv.12433}.
\bibitem[{Fontaine et~al.(2011)Fontaine, Paulo, Guimar\~aes, Kéfi, Loeuille,
  Memmott, Van~der Putten, Van~Veen and Thébault}]{Fontaineetal2011}
\bibinfo{author}{Fontaine, C.}, \bibinfo{author}{Paulo, R.},
  \bibinfo{author}{Guimar\~aes, J.}, \bibinfo{author}{Kéfi, S.},
  \bibinfo{author}{Loeuille, N.}, \bibinfo{author}{Memmott, J.},
  \bibinfo{author}{Van~der Putten, W.H.}, \bibinfo{author}{Van~Veen, F.J.F.},
  \bibinfo{author}{Thébault, E.}, \bibinfo{year}{2011}.
\newblock \bibinfo{title}{The ecological and evolutionary implications of
  merging different types of networks}.
\newblock \bibinfo{journal}{Ecology letters} \bibinfo{volume}{14},
  \bibinfo{pages}{1170--1181}.
\newblock \DOIprefix\doi{10.1111/j.1461-0248.2011.01688.x}.
\bibitem[{Girvan and Newman(2002)}]{Girvan2002}
\bibinfo{author}{Girvan, M.}, \bibinfo{author}{Newman, M.},
  \bibinfo{year}{2002}.
\newblock \bibinfo{title}{Community structure in social and biological
  networks}.
\newblock \bibinfo{journal}{Proc. Natl. Acad. Sci.} \bibinfo{volume}{99 (12)},
  \bibinfo{pages}{7821--7826}.
\newblock \DOIprefix\doi{10.1073/pnas.122653799}.
\bibitem[{Huaylla et~al.(2021)Huaylla, Nacif, Coulin, Kuperman and
  Garibaldi}]{Huayllaetal2021}
\bibinfo{author}{Huaylla, C.}, \bibinfo{author}{Nacif, M.},
  \bibinfo{author}{Coulin, C.}, \bibinfo{author}{Kuperman, M.},
  \bibinfo{author}{Garibaldi, L.}, \bibinfo{year}{2021}.
\newblock \bibinfo{title}{Decoding information in multilayer ecological
  networks: The keystone species case}.
\newblock \bibinfo{journal}{Ecological Modelling} \bibinfo{volume}{460},
  \bibinfo{pages}{109734}.
\newblock \DOIprefix\doi{https://doi.org/10.1016/j.ecolmodel.2021.109734}.
\bibitem[{Kuperman and Abramson(2001)}]{kuperman2001}
\bibinfo{author}{Kuperman, M.}, \bibinfo{author}{Abramson, G.},
  \bibinfo{year}{2001}.
\newblock \bibinfo{title}{Small world effect in an epidemiological model}.
\newblock \bibinfo{journal}{Phys. Rev. Lett.} \bibinfo{volume}{86},
  \bibinfo{pages}{2909--2912}.
\newblock \DOIprefix\doi{10.1103/PhysRevLett.86.2909}.
\bibitem[{Levin and Paine(1974)}]{Levin1974}
\bibinfo{author}{Levin, S.}, \bibinfo{author}{Paine, R.T.},
  \bibinfo{year}{1974}.
\newblock \bibinfo{title}{Disturbance, patch formation, and community
  structure.}
\newblock \bibinfo{journal}{Proc. Natl. Acad. Sci.} \bibinfo{volume}{71 (7)},
  \bibinfo{pages}{2744--2747}.
\newblock \DOIprefix\doi{10.1073/pnas.71.7.2744}.
\bibitem[{Meli\'an et~al.(2009)Meli\'an, Bascompte, Jordano and
  Krivan}]{Melian2009}
\bibinfo{author}{Meli\'an, C.}, \bibinfo{author}{Bascompte, J.},
  \bibinfo{author}{Jordano, P.}, \bibinfo{author}{Krivan, V.},
  \bibinfo{year}{2009}.
\newblock \bibinfo{title}{Diversity in a complex ecological network with two
  interaction types.}
\newblock \bibinfo{journal}{Oikos} \bibinfo{volume}{118 (1)},
  \bibinfo{pages}{122--130}.
\newblock \DOIprefix\doi{10.1111/j.1600-0706.2008.16751.x}.
\bibitem[{Morrison et~al.(2020)Morrison, Brosi and Dirzo}]{MorrisonDirzo2020}
\bibinfo{author}{Morrison, B.M.L.}, \bibinfo{author}{Brosi, B.J.},
  \bibinfo{author}{Dirzo, R.}, \bibinfo{year}{2020}.
\newblock \bibinfo{title}{Agricultural intensification drives changes in hybrid
  network robustness by modifying network structure}.
\newblock \bibinfo{journal}{Ecol. Lett.} \bibinfo{volume}{23},
  \bibinfo{pages}{359--369}.
\newblock \DOIprefix\doi{10.1111/ele.13440}.
\bibitem[{Morzy et~al.(2017)Morzy, Kajdanowicz and Kazienko}]{Morzy2017}
\bibinfo{author}{Morzy, M.}, \bibinfo{author}{Kajdanowicz, T.},
  \bibinfo{author}{Kazienko, P.}, \bibinfo{year}{2017}.
\newblock \bibinfo{title}{On measuring the complexity of networks: Kolmogorov
  complexity versus entropy}.
\newblock \bibinfo{journal}{Complexity} \bibinfo{volume}{2017}.
\newblock \DOIprefix\doi{https://doi.org/10.1155/2017/3250301}.
\bibitem[{Newman(2004a)}]{Newman2004a}
\bibinfo{author}{Newman, M.E.J.}, \bibinfo{year}{2004a}.
\newblock \bibinfo{title}{Analysis of weighted networks}.
\newblock \bibinfo{journal}{Phys. Rev. E.} \bibinfo{volume}{70},
  \bibinfo{pages}{056131--056140}.
\newblock \DOIprefix\doi{10.1103/PhysRevE.70.056131}.
\bibitem[{Newman(2004b)}]{Newman2004b}
\bibinfo{author}{Newman, M.E.J.}, \bibinfo{year}{2004b}.
\newblock \bibinfo{title}{Fast algorithm for detecting community structure in
  networks}.
\newblock \bibinfo{journal}{Phys. Rev. E.} \bibinfo{volume}{69 (6)},
  \bibinfo{pages}{066133}.
\newblock \DOIprefix\doi{10.1103/PhysRevE.69.066133}.
\bibitem[{Pilosof et~al.(2017)Pilosof, Porter, Pascual and
  K\'efi}]{Pilosofetal2017}
\bibinfo{author}{Pilosof, S.}, \bibinfo{author}{Porter, M.},
  \bibinfo{author}{Pascual, M.}, \bibinfo{author}{K\'efi, S.},
  \bibinfo{year}{2017}.
\newblock \bibinfo{title}{The multilayer nature of ecological networks}.
\newblock \bibinfo{journal}{Nat Ecol Evol} \bibinfo{volume}{1},
  \bibinfo{pages}{0101}.
\newblock \DOIprefix\doi{https://doi.org/10.1038/s41559-017-0101}.
\bibitem[{Poulin(2010)}]{Poulin2010}
\bibinfo{author}{Poulin, R.}, \bibinfo{year}{2010}.
\newblock \bibinfo{title}{Parasite manipulation of host behavior: An update and
  frequently asked questions}.
\newblock \bibinfo{journal}{Elsevier} \bibinfo{volume}{41},
  \bibinfo{pages}{151--186}.
\bibitem[{Solomonoff(1964)}]{Solomonoff19641}
\bibinfo{author}{Solomonoff, R.}, \bibinfo{year}{1964}.
\newblock \bibinfo{title}{A formal theory of inductive inference. part i}.
\newblock \bibinfo{journal}{Information and Control} \bibinfo{volume}{7},
  \bibinfo{pages}{1--22}.
\newblock \DOIprefix\doi{https://doi.org/10.1016/S0019-9958(64)90223-2}.
\bibitem[{Strydom et~al.(2021)Strydom, Dalla~Riva and Poisot}]{Strydom2021}
\bibinfo{author}{Strydom, T.}, \bibinfo{author}{Dalla~Riva, G.V.},
  \bibinfo{author}{Poisot, T.}, \bibinfo{year}{2021}.
\newblock \bibinfo{title}{Svd entropy reveals the high complexity of ecological
  networks}.
\newblock \bibinfo{journal}{Front. Ecol. Evol.} \bibinfo{volume}{9},
  \bibinfo{pages}{623141}.
\newblock \DOIprefix\doi{10.3389/fevo.2021.623141}.
\bibitem[{Watts and Strogatz(1998)}]{Watts1998}
\bibinfo{author}{Watts, D.J.}, \bibinfo{author}{Strogatz, S.H.},
  \bibinfo{year}{1998}.
\newblock \bibinfo{title}{Collective dynamics of 'small-world' networks}.
\newblock \bibinfo{journal}{Nature} \bibinfo{volume}{393},
  \bibinfo{pages}{440--442}.
\newblock \DOIprefix\doi{https://doi.org/10.1038/30918}.
\bibitem[{Zenil et~al.(2016)Zenil, Hernández-Orozco, Kiani, Soler-Toscano and
  Rueda-Toicen}]{Zeniletal2016}
\bibinfo{author}{Zenil, H.}, \bibinfo{author}{Hernández-Orozco, S.},
  \bibinfo{author}{Kiani, N.}, \bibinfo{author}{Soler-Toscano, F.},
  \bibinfo{author}{Rueda-Toicen, A.}, \bibinfo{year}{2016}.
\newblock \bibinfo{title}{A decomposition method for global evaluation of
  shannon entropy and local estimations of algorithmic complexity}.
\newblock \bibinfo{journal}{Entropy} \bibinfo{volume}{20},
  \bibinfo{pages}{605}.
\newblock \DOIprefix\doi{https://doi.org/10.3390/e20080605}.

\end{thebibliography}





\end{document}


\maketitle

In this research, we used 23 networks of plant-pollinator and 19 networks of host-parasite interactions from the web-of-life.es database.

The nestedness and SVD entropy values of the selected networks were calculated and plotted in the following figures.

\begin{figure}[H]
\centering
\includegraphics[scale=0.40]{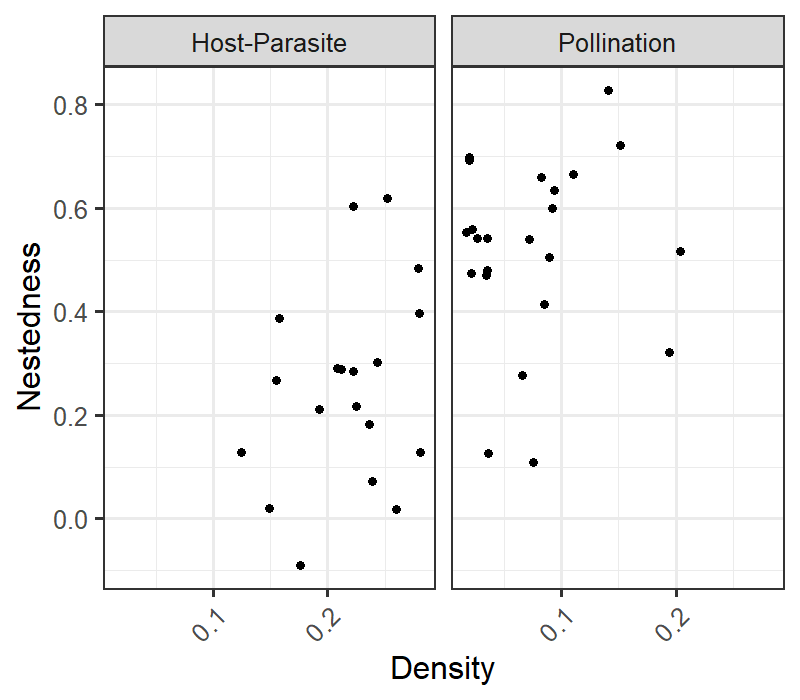}
\caption{Relation between nestedness and density from Host-Parasite network (left) and Plant-Pollinator network (right)}
\label{anidamiento}
\end{figure}

\begin{figure}[H]
\centering
\includegraphics[scale=0.40]{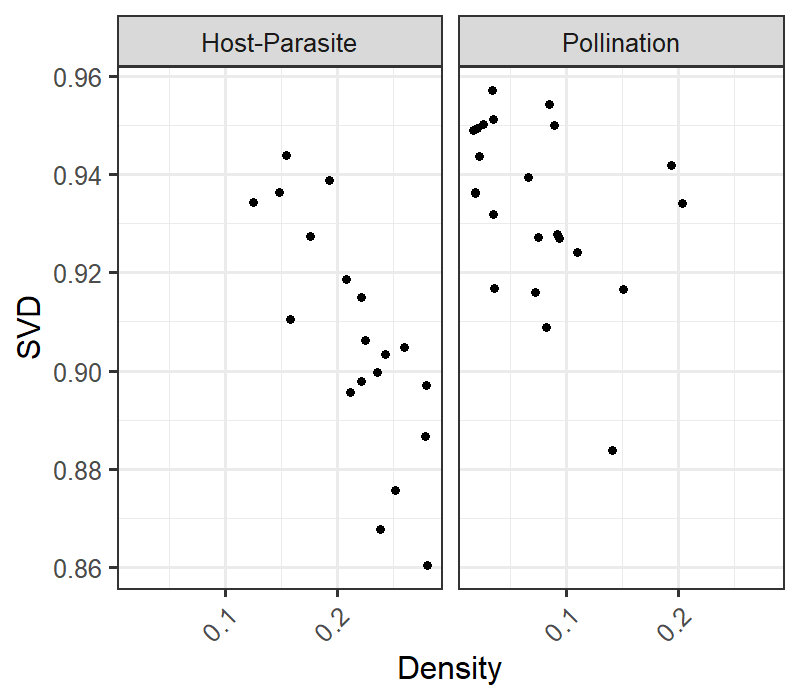}
\caption{SVD entropy vs density using original values in host-parasite network (left) and plant-pollinator network (right)}
\label{svdoriginal}
\end{figure}

The tables (A.1, A.2) show the name of the network, which can be found on the website, the number of nodes, links, density, and the study site where the data were collected.

\begin{table}[H]
\caption{List of selected host-parasite networks to be studied}
\begin{center}
\resizebox{15cm}{!}{
\large
\begin{tabular}{p{4cm} p{2.5cm}p{2.5cm}p{2.5cm}p{6cm}}
\hline
Network & Nodes	& Links & Density &	Site \\ \cline{1-5}
\hline
\textit{A\_HP\_002} &	42 & 96 & 0.222 & Akmolinsk\\ \hline
\textit{A\_HP\_006} & 53 & 123 &	0.208 & Armenia\\ \hline 
\textit{A\_HP\_008} &  32 & 37 & 0.193 & Chimkent\\ \hline
\textit{A\_HP\_010} &	49 & 88 & 0.158 &  East Balkhash\\ \hline
\textit{A\_HP\_019} & 27 & 36	& 0.222 & Krasnojarsk\\ \hline
\textit{A\_HP\_022} & 34 & 68	& 0.236 & Kustanai\\ \hline
\textit{A\_HP\_024} & 27 & 39	& 0.279 & Mongolia,Central Khangay\\ \hline
\textit{A\_HP\_025} & 58 & 107 & 0.149 & Mongolia,North Western Khangay\\ \hline
\textit{A\_HP\_027} & 47 & 108 &	0.212 &	 Moyyunkum\\ \hline
\textit{A\_HP\_029} & 49 & 79	& 0.155 & Kyrgyz Republic\\ \hline
\textit{A\_HP\_030} & 29 & 51 & 0.243 & North Russian Far East\\ \hline
\textit{A\_HP\_031} & 56 & 134 &	0.225 &	Novosibirsk\\ \hline
\textit{A\_HP\_032} & 27 & 32 & 0.176 & Pavlodar\\ \hline
\textit{A\_HP\_037} & 38 & 90 & 0.252 & Slovakia\\ \hline
\textit{A\_HP\_038} & 30 & 52 & 0.26 & Southwestern Azerbajan\\ \hline
\textit{A\_HP\_042} & 53 & 84 & 0.125 & Tarbagatai\\ \hline
\textit{A\_HP\_043} & 38 & 73	& 0.28 & Terskey Alatau\\ \hline
\textit{A\_HP\_044} & 27 & 197 & 0.281 & Tomsk, Tumen\\ \hline
\textit{A\_HP\_050} & 62 & 226 &	0.239 & Volga, Kama\\ \hline
\end{tabular}
}
\end{center}
\label{hp}
\end{table}

\begin{table}[H]
\caption{List of selected Pollinator-plant networks to be studied}
\begin{center}
\resizebox{15cm}{!}{
\large
\begin{tabular}{p{4cm} p{2.5cm}p{2.5cm}p{2.5cm}p{6cm}}
\hline
Network & Nodes	& Links & Density &	Site \\ \cline{1-5}
\hline
\textit{M\_PL\_005} &	371 &	923	& 0.035	& Pikes Peak, Colorado, USA\\ \hline
\textit{M\_PL\_006} &	78 & 146 & 0.141	& Hickling, Norfolk, UK\\ \hline
\textit{M\_PL\_009} & 142 & 242 & 0.085 & Latnjajaure, Abisko, Sweden \\ \hline
\textit{M\_PL\_010} & 107 &	456 & 0.194 &	Zackenberg \\ \hline
\textit{M\_PL\_013} & 65 & 103 & 0.204 & KwaZulu, Natal region, South Africa\\ \hline
\textit{M\_PL\_015} & 797 &	2933 &	0,034 &	Daphn\'i, Athens, Greece \\ \hline
\textit{M\_PL\_016} & 205 & 412 & 0.089 & Do\~nana Nat. Park, Spain \\ \hline
\textit{M\_PL\_017} & 104 &	299 &	0.151 &	Bristol, England \\ \hline
\textit{M\_PL\_021} & 768 &	1193 & 0.019 &	Ashu, Kyoto, Japan \\ \hline
\textit{M\_PL\_023} & 95 & 125 & 0.075 & Rio Blanco, Mendoza, Argentina \\ \hline
\textit{M\_PL\_026} & 159 &	204 &	0.036 &	Galapagos \\ \hline
\textit{M\_PL\_031} & 97 &	156 & 0.066 &	Canaima Nat. Park, Venezuela \\ \hline
\textit{M\_PL\_034} & 154 &	312 & 0.094 & Chiloe, Chile \\ \hline
\textit{M\_PL\_044} & 719 &	1125 &	0.017 &	Amami, Ohsima Island, Japan \\ \hline
\textit{M\_PL\_047} & 205 &	425 & 0.11	& Isenbjerg \\ \hline 
\textit{M\_PL\_048} &	266 & 671 &	0.092 &	Denmark \\ \hline
\textit{M\_PL\_053} & 393 &	589 &	0.021 &	Mt. Yufu, Japan \\ \hline
\textit{M\_PL\_054} &	431 & 773 &	0.022 &	Kyoto City, Japan \\ \hline
\textit{M\_PL\_055} &	259 &	431 & 0.035 &	Nakaikemi marsh, Fukui Prefecture, Japan \\ \hline
\textit{M\_PL\_056} & 456 &	871 &	0.026 &	Mt. Kushigata, Yamanashi Pref, Japan \\ \hline
\textit{M\_PL\_057} & 997 &	1920 & 0.019 &	Kibune, Kyoto, Japan \\ \hline
\textit{M\_PL\_072\_04} & 94 &	192 &	0.072 &	Difuntos, Pampas, Argentina \\ \hline
\textit{M\_PL\_072\_05} & 84 & 138 & 0.082 &	El Morro, Pampas, Argentina \\ \hline
\end{tabular}
}
\end{center}
\label{poli}
\end{table}